\documentstyle[psfig,mncite]{mn}

\newcommand{\lta}{{\small\raisebox{-0.6ex}
{$\,\stackrel{\raisebox{-.2ex}{$\textstyle <$}}{\sim}\,$}}}

\title[The secondary star in CVs and LMXBs]{The secondary star in cataclysmic variables 
and low mass X-ray binaries}

\author[D.\,A.\ Smith and V.\,S.\ Dhillon]{D.\,A.\ Smith,$^{1,2}$
V.\,S.\ Dhillon$^{3}$\\
$^1$Institute of Astronomy, Madingley Road, Cambridge CB3 0HA\\
$^2$Royal Greenwich Observatory, Madingley Road, Cambridge CB3 0EZ\\
$^3$Department of Physics and Astronomy, University of Sheffield, Sheffield S3 7RH\\
email: dsmith@ast.cam.ac.uk, vik.dhillon@sheffield.ac.uk}

\date{\center{\Large Accepted For Publication in MNRAS. \\
\vspace{.5cm} \today}}

\begin{document} 
\maketitle
\begin{abstract}
We critically re-examine the available data on the spectral types, masses and radii 
of the secondary stars in cataclysmic variables (CVs) and low-mass X-ray binaries 
(LMXBs), using the new catalogue of \protect\scite{ritter98} as a starting point. 
We find there are 55 reliable spectral type determinations and only 14 reliable 
mass determinations of CV secondary stars (10 and 5, respectively, in the case of 
LMXBs). We derive new spectral type--period, mass--radius, mass--period and
radius--period relations, and compare them with theoretical predictions. We find 
that CV secondary stars with orbital periods shorter than 7--8 hours are, as a group, 
indistinguishable from main sequence stars in detached binaries. We find it is not 
valid, however, to estimate the mass from the spectral type of the secondary star in 
CVs or LMXBs. We find that LMXB secondary stars show some evidence for evolution, with 
secondary stars which are slightly too large for their mass. We show how the masses 
and radii of the secondary stars in CVs can be used to test the validity of the 
disrupted magnetic braking model of CV evolution, but we find that the currently 
available data are not sufficiently accurate or numerous to allow such an analysis. 
As well as considering secondary star masses, we also discuss the masses of the 
white dwarfs in CVs, and find mean values of ${\overline M}_1 = 0.69\pm0.13M_\odot$ 
below the period gap, and ${\overline M}_1 = 0.80\pm0.22M_\odot$ above the period gap.
\end{abstract} 

\begin{keywords}
binaries: close -- novae, cataclysmic variables -- X-rays: stars -- late type stars.
\end{keywords}

%

\section{Introduction}
\label{sec:introduction}
Cataclysmic variables (CVs) are semi-detached binary stars in which a white dwarf primary 
accretes material from a Roche-lobe filling secondary. For a thorough review of CVs see 
\scite{warner95a}. The spectral type and luminosity class of the secondary star can be 
estimated from basic theory, as follows. Kepler's third law can be written as
\begin{equation}
{{4\pi^2 a^3} \over {GP^2}} = {M_1+M_2}={M_2{\Bigl({{1+q} \over q}\Bigr)}},
\end{equation}
where $q = M_2/M_1$. The volume-equivalent radius of the Roche lobe can be 
approximated by
\begin{equation}
{{R_2} \over a} = 0.47{\Bigl({q\over{1+q}}\Bigr)}^{1/3}.
\end{equation}
This relation is a slightly modified form of the \scite{paczynski71} equation, and
we have found that it is accurate to less than 3 per cent over the range of mass ratios relevant 
for CVs ($0.01<q<1.0$). Combining equations (1) and (2) gives the mean density--period 
relation,
\begin{equation}
{{\rho} \over {\rho_\odot}} = \Bigl({M_2 \over M_\odot}\Bigr)\Bigl({R_2 \over R_\odot}\Bigr)^{-3} = 75.5 P^{-2}({\rm hr}),
\end{equation}
which is accurate to $\sim6$ per cent (see \pcite{eggleton83}). 

Typical lower main-sequence mean densities range from $\sim50\rho_\odot$ for M8 dwarfs \cite{allen76}, 
corresponding to the minimum orbital period of CVs around 80\,min, to $\sim1\rho_\odot$ for G0 dwarfs, 
corresponding to an orbital period of around 9\,hr. At longer periods than $\sim9$\,hr, the density of 
the secondary star must be sub-solar, corresponding to F-type (and earlier) main sequence stars.
F-type main sequence stars have masses above solar ($M \sim 1.3M_\odot$ for an F5 dwarf, 
$M \sim1.6M_\odot$ for an F0 dwarf, \pcite{gray92}), which is relevant when considering the formal 
requirement of $q<5/6$ for stable, conservative mass transfer (e.g. \pcite{frank92}). Since the 
mass of the white dwarf must be below the Chandrasekhar limit, the secondary is forced to have 
$M_2 < 1.2M_\odot$. As the white dwarf population is biased towards much lower masses (the mean 
white dwarf mass in CVs is $0.77 \pm 0.21 M_\odot$, see section~6), the number of systems with 
secondary stars above solar mass is very small indeed. Secondary stars with sub-solar mean densities 
in long period CVs must therefore be low mass evolved M or K-stars rather than intermediate 
mass main sequence F-stars. In summary, for the orbital period range in which most CVs lie 
($1.3$\,hr$ \lta P \lta 9$\,hr), the secondary stars should be M, K or G main-sequence dwarfs, 
while longer period systems must harbour secondaries which have evolved away from the main sequence. 

Even though we have just shown that most CV secondaries should have lower main-sequence 
densities, it is not clear that they should appear as lower main-sequence stars. This 
is because CV secondaries are subject to a number of extreme environmental factors to 
which field stars are not. Specifically, CV secondaries are:
\begin{enumerate}
\item 
situated $\sim1R_\odot$ from a hot, irradiating source (see \pcite{smith95}),
\item 
rapidly rotating ($\sim 100$~km\,s$^{-1}$),
\item 
Roche-lobe shaped,
\item 
losing mass at a rate of $\sim10^{-8} - 10^{-11}M_\odot$\,yr$^{-1}$,
\item 
survivors of a common-envelope phase during which they existed within the atmosphere 
of a giant star,
\item 
exposed to nova outbursts every $\sim10^4$\,yr. 
\end{enumerate}

It is the purpose of this paper to see if the above environmental factors alter the 
gross properties (masses, radii, spectral types) of the secondary stars in CVs. This 
is not the first time that the question of whether CV secondaries lie on the main 
sequence has been investigated. \scite{echevarria83} concluded that they were not. 
He came to this conclusion by calculating an empirical mass--radius relation for 
field stars and then using a density--period relation similar to equation~(3) to 
calculate a mass--period relation. This was then combined with a spectral type--mass 
relation to get a spectral type--period relation for field stars. By comparing this 
relation with the measured spectral types and periods of 17 CVs he found that the CVs 
did not fit his relation and concluded that they were therefore not on the main sequence. 

Warner (1995a,b) disputed Echevarr\'{\i}a's conclusion by arguing that he had equated 
a poorly-fitting power law to the mass--spectral type relationship and that as a 
result of this Echevarr\'{\i}a's derived relationship systematically predicted 
spectral types that were too early for long period CVs and too late for the short period 
CVs. Instead, Warner plotted the CV secondary stars on a spectral type--period diagram
and calculated equivalent periods of field stars using a mass--period relation,
derived using a mass--radius relation which fits both the CV data of \scite{webbink90} 
and the field star data of \scite{popper80}.
Warner concluded that, as a group, secondary stars have masses, radii and spectral 
types related in exactly the same way as main sequence stars, although some 
individual CVs do depart from the average properties. A similar conclusion has been 
reached by \scite{ritter83} and \scite{patterson84}.

Low mass X-ray binaries (LMXBs) are also semi-detached systems, the difference being that 
the primary in these systems is a neutron star or a black hole. The secondary stars in LMXBs 
should be similar to those in CVs, except that they are irradiated to a much greater degree by X-rays. 
Also, the formation of LMXBs containing neutron stars requires that the secondary stars have 
a mass between $1.3-1.5M_\odot$ at the onset of mass transfer, and that they should be 
significantly nuclear evolved \cite{king97}. LMXB secondaries should therefore appear even 
less like main sequence stars than their CV counterparts.

In this paper we repeat the analyses of \scite{echevarria83}, \scite{ritter83}, \scite{patterson84}, 
\scite{webbink90} and Warner (1995a,b) and go a few steps further, using the more 
extensive data set now available (\pcite{ritter98} and other sources) to derive the 
first reliable spectral type--period relation for CVs. We present new mass--radius, 
mass--period and radius--period relations for CV secondary stars with reliable system 
parameters, and compare the parameters of the CV and LMXB secondaries with those of detached 
binaries as listed in the reviews of \scite{andersen91} and \scite{popper80}. We also 
briefly examine the observed mass distribution of the white dwarf population in CVs, and conclude 
with a discussion on how the measured masses and radii of CV secondary stars can be used to
constrain the disrupted magnetic braking model of CV evolution.

%

\section{The Orbital Period--Spectral Type relation}
\label{sec:period}

\begin{figure*}
\label{fig:sp1}
\psfig{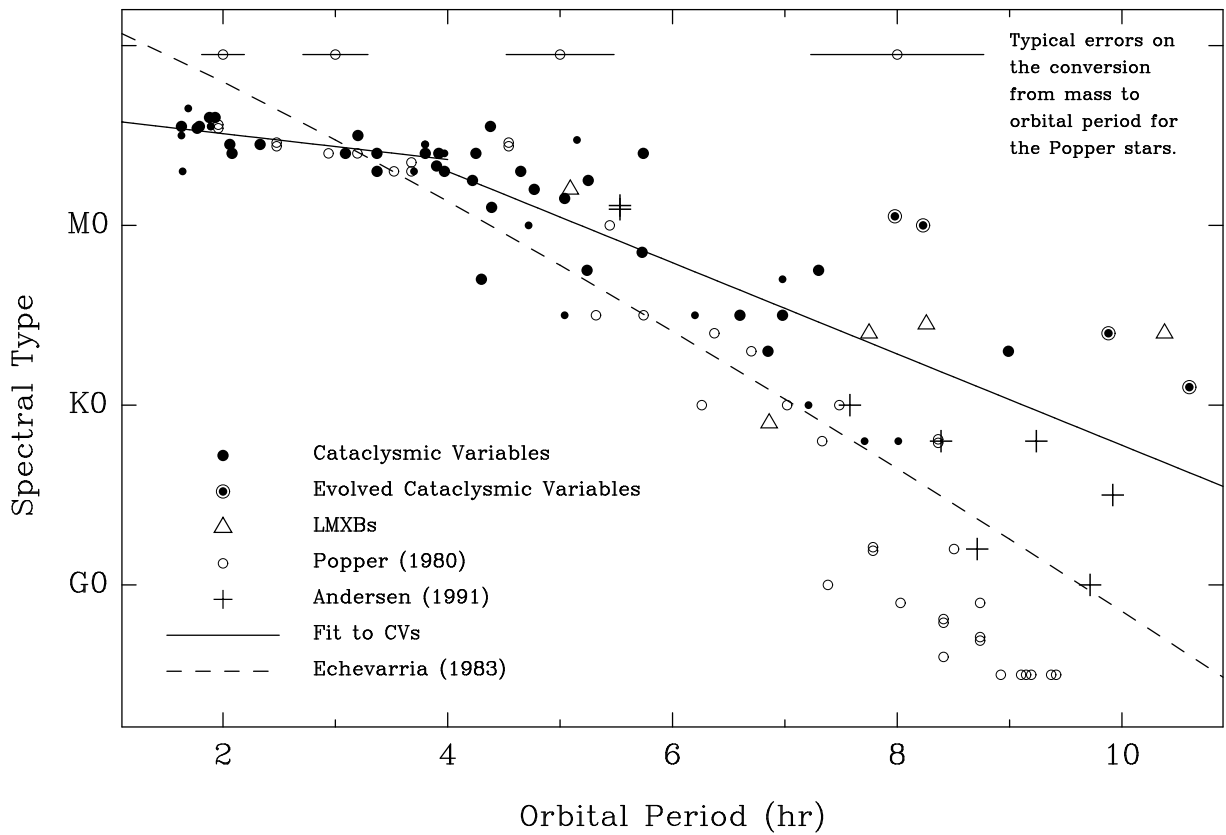}
\caption{The spectral types of the secondary stars in CVs and LMXBs versus their orbital 
periods. The CVs have been plotted as filled circles, large circles denoting better spectral 
type determinations (weight = 1 in Table~1) than the small circles (weight = 0.5). 
LMXBs are denoted by triangles. Also plotted are 50 isolated stars for which the 
spectral types and masses have been measured (see text for details). The solid line 
is a two part linear fit to the CVs, given by equation~(4). The dashed line is the 
relationship derived by Echevarr\'{\i}a (1983).}
\end{figure*}

The spectral type of the secondary star has been measured in 66 CVs with 
known orbital periods, according to the catalogue of \scite{ritter98}. A number of these
measurements, however, are dubious, relying on main sequence assumptions, or infrared
colours which are contaminated by disc emission \cite{berriman85}, to determine the 
spectral types. The only reliable method of determining spectral types is to detect 
absorption features in the spectra of the secondary stars. One can then compare the
equivalent widths of the secondary star absorption features with those of isolated dwarf 
stars (e.g. \pcite{wade88}), or simply match the observed absorption features with those 
in a set of template spectra, either by eye (e.g. Friend 1990a,b) or by using an optimal 
subtraction technique (e.g. \pcite{smith98}). Absorption features have been detected 
(and hence reliable spectral types determined) in 55 of the 66 CVs catalogued by 
\scite{ritter98}, as listed in Table~1. Most of the spectral types listed in Table~1 do 
not have error bars associated with them, so we have instead assigned a weighting to each 
determination. In most cases, a weighting of 1 implies a careful spectral type determination 
using ample template dwarfs and a weighting of 0.5 implies there is a larger uncertainty in 
the exact value of the spectral type, often as a result of insufficient template dwarfs 
having been observed. 

The spectral types of the secondary stars listed in Table~1 are plotted against orbital 
period in Fig.~1. Two linear least-squares fits were performed, one above and one below 
the kink at an orbital period of 4~hours. Note that we omitted 7 CVs from the fit (indicated 
by the asterisks in Table~1) which are believed to harbour evolved secondary stars. 
For the sake of simplicity we have treated each spectral type as being divided into ten 
equal sub-types.\footnote{The MK system officially does not contain the K8 or K9 subtypes, nor those of 
M4.5, M5.5 etc. \protect\cite{jaschek87}. However these are in (infrequent) use, so to 
make the spectral type--period relation simple to use, and since the system is in any case 
non-linear, we have adopted a system where each spectral type is divided into 10 subtypes. 
This system was also followed by \protect\scite{echevarria83} and \protect\scite{warner95b} 
in their spectral type--period diagrams.} The resulting fits are given by:

\begin{equation}
\begin{array}{rrrll}
Sp(2)\,\,\, = & \! \,\,26.5 & \!\! - \,\, 0.7 & \!\!\!\!\! P, & P < 4{\rm hr}\\
 &  \pm \,\, 0.7 & \!\! \pm \,\, 0.2 & \\
 & & & & \\
 = & \! \,\,33.2 & \!\! - \,\, 2.5 & \!\!\!\!\! P, & P > 4{\rm hr}\\
 &  \pm \,\, 3.1 & \!\! \pm \,\, 0.5 & \\
\end{array}
\end{equation}
where $Sp(2)=0$ represents a spectral type G0, $Sp(2)=10$ is K0 and $Sp(2)=20$ is M0.
The rms scatter is 0.8 spectral sub-types for $P < 4$\,hr, and 3.0 sub-types for $P > 4$\,hr.  

The solid lines in the spectral type--period diagram represent our two-part linear 
fit to the CVs (equation~4), while the dashed line represents Echevarr\'{\i}a's equation.
Also plotted as a comparison in Fig.~1 are the Sun and 49 other late-type main sequence 
stars in detached binaries. Twelve of these 49 have been taken from the list of 
\scite{andersen91}, while the review paper by \scite{popper80} lists a further 37 stars with 
less well refined mass and radius measurements, but which are suitable for comparison purposes; 
we have taken 11 stars from Popper's Table~2 (Detached main sequence binaries, B6 to M), 
6 stars from Table~7 (Resolved spectroscopic binaries), and 20 from Table~8 (Visual binaries). 

The stars in detached binaries are plotted in Fig.~1, not according to their actual orbital 
periods, but rather according to the period of a CV containing a Roche lobe-filling secondary 
star of that mass and radius. This assumes, therefore, that the spectral types of the stars in 
detached binaries would remain unchanged if they were to become lobe-filling secondaries. There 
are two ways of performing the conversion from mass and radius to period, depending on how 
accurately the radii of the detached stars are known. In the case of the stars listed by Popper, 
which have poorly determined radii, we have converted from mass to period using the empirical 
mass--period relation we derive in section~4 (equation~8). Typical errors in this conversion,
as calculated from the errors in equation~(8), are plotted for different orbital periods at the 
top of Fig.~1. In the case of the stars listed by Andersen, all of which are eclipsing binaries 
and hence have accurately determined radii, we have used equation~(3) directly to convert from 
mass and radius to period. These have a much smaller associated error (3 per cent), so the 
crosses in Fig.~1 (Andersen's points) should be given greater weighting than the open circles 
(Popper's points) when interpreting the diagram.

There appears to be little difference between the distributions of the CVs and the main 
sequence stars in Fig.~1 for periods below 4~hours. Above 4~hours, there is much larger 
scatter in the two distributions but they still seem to agree up to 7--8~hours. Beyond 
7--8~hours the two distributions show a definite divergence and some evolved secondary 
stars appear, while beyond 9~hours the secondary stars all show signs of evolution (as 
is to be expected, see section~1). Echevarr\'{\i}a's relationship is a poor fit to the 
detached systems, as noted by Warner (1995a,b), and can be seen to erroneously predict 
later spectral types at short periods, and earlier spectral types at long periods, than 
those which have been observed. Note that the Echevarr\'{\i}a relationship was used as 
the basis for the claim by Friend et al. (1990a,b) that the secondary stars in CVs were 
``too cool for comfort'' and ``too cool for credibility,'' which no longer appears to 
be true (but see Section~4).

In Fig.~1, we have also plotted the 5 low-mass X-ray binaries (LMXBs) with known orbital 
periods (below 12 hours) and spectral types. They are listed 
in Table~2. To within the scatter, the LMXBs appear to have the same distribution as the 
CVs, indicating that LMXBs with orbital periods of 7--8 hours and longer harbour evolved 
secondary stars.

\begin{table*}
\label{tab:sp}
\caption{Spectral types of secondary stars and orbital periods (in hours) of cataclysmic 
variables. Key: CN = classical nova; DN = dwarf nova; RN = recurrent nova; NL = nova-like; 
P = polar; IP = intermediate polar; * = systems with "evolved" secondary stars, omitted 
from the linear fit in Fig.~1.}
\begin{tabular}{lcccccclccccc} 
\hline
\multicolumn{1}{l}{Star} &
\multicolumn{1}{c}{Class} & 
\multicolumn{1}{c}{$P_{orb}$} & 
\multicolumn{1}{c}{Sp Type} &
\multicolumn{1}{c}{Weight} &
\multicolumn{1}{c}{Ref.} &
\multicolumn{1}{l}{ } &
\multicolumn{1}{l}{Star} &
\multicolumn{1}{c}{Class} & 
\multicolumn{1}{c}{$P_{orb}$} & 
\multicolumn{1}{c}{Sp Type} &
\multicolumn{1}{c}{Weight} &
\multicolumn{1}{c}{Ref.} \\
\hline
BZ UMa	 &	DN  & 	1.63 &	M5.5V 	& 1 	& 4,5	& \, &	DQ Her	&	CN,IP &	4.65 &	M3$^+$V  & 1 	& 42,43 \\
RX J0719+655 & 	P   & 	1.63 &  M4-6V 	& 0.5 	& 1	& \, &	UX UMa	&	NL  &	4.72 &	M0V      & 0.5 	& 44,45 \\
EX Hya	 &	IP  &	1.64 &	M3V    	& 0.5	& 6,7 	& \, &	V895 Cen &  	P   & 	4.77 & 	M2V	& 1	& 74	\\
V834 Cen &  	P   &	1.69 &	M6.5V  	& 0.5 	& 8,9 	& \, &  EX Dra	&	DN  &	5.04 &	M1-2V  	& 1 	& 46,47 \\
HT Cas	 &	DN  &	1.77 &	M5.4V	& 1	& 10,11 & \, &  RX And 	&	DN  &	5.04 &	K5V  	& 0.5 	& 48,49	\\
Z Cha	 &	DN  &	1.79 &	M5.5V	& 1 	& 12,13 & \, &  AR Cnc 	&	DN  &	5.15 &	M4-5.5V  & 0.5 	& 2,3 	\\
V2301 Oph &	P   &	1.88 &	M6V  	& 1 	& 14,15 & \, &  EY Cyg	&	DN  &	5.24 &	K5-M0V	& 1 	& 50,28 \\
MR Ser	 &    	P   &	1.89 &	M5-6V   & 0.5 	& 16,17 &  \, & CZ Ori	&	DN  &	5.25 &	M2.5V	& 1 	& 4 	\\
ST LMi	 &  	P   &	1.89 &	M5-6V  	& 0.5 	& 17,18 &  \, & AT Cnc	&	DN  &	5.73 &	K7-M0V	& 1 	& 51,28 \\
AR UMa	 &	P   &	1.93 &	M6V	& 1 	& 19	& \, &	AH Eri	&	DN  &	5.74 &	M3-5V	& 1 	& 72,73 \\
DV UMa	 &	DN  & 	2.06 &	M4-5V	& 1 	& 20,3 	& \, &	AH Her 	&	DN  &	6.20 &	K5V   	& 0.5	& 52,53 \\
HU Aqr   &  	P   &	2.08 &	M4V  	& 1 	& 21  	& \, & 	SS Cyg 	&	DN  &	6.60 &	K5V    	& 1 	& 54	\\
QS Tel	 &	P   &	2.33 &	M4.5V	& 1 	& 22	& \, &	V426 Oph &	DN  &	6.85 &	K3V  	& 1 	& 55   	\\
AM Her	 &  	P   & 	3.09 &	M4$^+$V	& 1 	& 23  	& \, &  Z Cam 	&	DN  & 	6.98 &	K7V      & 0.5 	& 56,57 \\
MV Lyr	 &	NL  &	3.20 &	M5V  	& 1 	& 24,25 & \, &  EM Cyg 	&	DN  &	6.98 &	K5V    	& 1 	& 58,59 \\
V1432 Aql &  	P   &	3.37 &	M4V     & 1 	& 26 	& \, &	AC Cnc 	&	NL  &	7.21 &	K0V   	& 0.5 	& 60,61 \\
UU Aql	 &	DN  &	3.37 &	M2-4V	& 1 	& 27,28 & \, &	TT Crt 	&  	DN  &	7.30 &	K5-M0V 	& 0.5 	& 62 	\\
QQ Vul	 &  	P   &	3.71 &	M2-4V	& 0.5 	& 29,17 & \, &  V363 Aur$^2$ &	NL  &	7.71 &	G8V  & 0.5 	& 63 	\\
IP Peg 	&	DN  &	3.80 &	M4.5V	& 0.5 	& 30,31 & \, &  V1309 Ori* &	P   & 	7.98 &	M0-1V  	&  --	& 64,65 \\
VY For & 	P   &	3.80 &	M4.5V  	& 1 	& 32	& \, & 	BT Mon	&	CN  & 	8.01 & 	G8V    	& 0.5 	& 66	\\
KT Per	&	DN  &	3.90 &	M3.3V	& 1 	& 33 	& \, & 	CH UMa*	&	DN  &	8.23 &	M0V 	&  --	& 54	\\
CN Ori	&	DN  &	3.92 &	M4$^+$V & 1 	& 34,35 & \, &	RU Peg  &	DN  &	8.99 &	K3V  	& 1 	& 67,54 \\
DO Dra 	&	DN,IP &	3.97 &	M4V     & 1 	& 36,80 & \, &	AE Aqr* &	IP  & 	9.88 &	K4V	&  --	& 68,69 \\
WW Cet	&	DN  & 	4.22 &	M2.5V	& 1 	& 37,38 & \, & 	DX And* &    	DN  &  10.6  &	K1V	&  --	& 70 	\\
U Gem 	&	DN  &	4.25 &	M4$^+$V & 1 	& 39,35 & \, &	U Sco*  &	RN  &  29.5  &  F8V 	&  --   & 75,76 \\
BD Pav	&	DN  &	4.31 &	K7V	& 1 	& 35 	& \, &	GK Per*	& CN,IP,DN  &  47.9  &  K2-3IV-V &  --  & 77,78\\
TW Vir$^1$ &	DN  &	4.38 &	M5-6V 	& 0.5 	& 40,28 & \, & V1017 Sgr*& CN,DN   &  137.1  &  G5IIIp 	&  -- 	& 79,71\\
SS Aur 	&	DN  &	4.39 &	M1V    	& 1 	& 41,35 & & & & & & \\
\hline
\end{tabular}
\begin{tabular}{l}
References:
1. \pcite{tovmassian97},
2. \pcite{howell90},
3. \pcite{mukai90},
4. \pcite{ringwald94},\\
\hspace{1.46cm}	
5. \pcite{jurcevic94},
6. \pcite{sterken83},
7. \pcite{dhillon97b},
8. \pcite{schwope93},	
9. \pcite{puchnarewicz90},\\
\hspace{1.31cm}	
10. \pcite{horne91b},
11. \pcite{marsh90},
12. \pcite{robinson95},
13. \pcite{wade88}, \\	
\hspace{1.31cm}	
14. \pcite{barwig94a},  
15. \pcite{silber94},
16. \pcite{schwope91},
17. \pcite{mukai86},\\
\hspace{1.31cm}		
18. \pcite{cropper86},
19. \pcite{remillard94},
20. \pcite{howell88},	
21. \pcite{glenn94},\\
\hspace{1.31cm}		
22. \pcite{schwope95},
23. \pcite{young81a},
24. \pcite{skillman95},\\
\hspace{1.31cm}		
25. \pcite{schneider81},
26. \pcite{watson95},
27. \pcite{ritter98},
28. \pcite{smith97a},\\
\hspace{1.31cm}	
29. \pcite{andronov87},
30. \pcite{wolf93},
31. \pcite{martin87},
32. \pcite{beuermann89a},\\
\hspace{1.31cm}	
33. \pcite{thorstensen97a},
34. \pcite{barrera89},
35. \pcite{friend90a},
36. \pcite{haswell97},\\
\hspace{1.31cm}	
37. \pcite{ringwald96},
38. \pcite{hawkins90},
39. \pcite{smak93},
40. \pcite{shafter83a},
41. \pcite{shafter86b},\\
\hspace{1.31cm}	
42. \pcite{zhang95},
43. \pcite{young81b},
44. \pcite{baptista95},
45. \pcite{rutten94b},
46. \pcite{fiedler94},\\
\hspace{1.31cm}		
47. \pcite{billington96},
48. \pcite{kaitchuck89},
49. \pcite{dhillon95},
50. \pcite{sarna95},\\
\hspace{1.31cm}	
51. \pcite{goetz86},
52. \pcite{horne86b},
53. \pcite{bruch87},	
54. \pcite{friend90b},
55. \pcite{hessman88},\\
\hspace{1.31cm}	
56. \pcite{thorstensen95},
57. \pcite{szkody81},
58. \pcite{stover81a},	\\
\hspace{1.31cm}	
59. \pcite{beuermann84},
60. \pcite{okazaki82},	
61. \pcite{schlegel84},\\
\hspace{1.31cm}		
62. \pcite{szkody92},	
63. \pcite{schlegel86},  
64. \pcite{buckley95},
65. \pcite{shafter95},\\
\hspace{1.31cm}	
66. \pcite{smith98},
67. \pcite{stover81b},
68. \pcite{welsh93},	
69. \pcite{casares96},\\
\hspace{1.31cm}			
70. \pcite{drew93}, 	
71. \pcite{sekiguchi92},
72. \pcite{thorstensen97b},
73. \pcite{howell94},\\
\hspace{1.31cm}
74. \pcite{stobie96},
75. \pcite{schaefer95},
76. \pcite{johnston92},
77. \pcite{reinsch94},\\
\hspace{1.31cm}
78. \pcite{crampton86},
79. \pcite{kraft64},
80. \pcite{mateo91}.\\
\end{tabular}
\begin{tabular}{l}
$^1$ Spectral type uncertain since it was determined from a spectrum taken 
while the star was on the rise to outburst. \\
$^2$ Spectral type given as late G.\\
\end{tabular}
\end{table*}

\begin{table*}
\label{tab:splmxb}
\caption{Spectral types, masses and radii of secondary stars in LMXBs.}
\begin{tabular}{lccccccc} 
\hline
\multicolumn{1}{l}{Star} &
\multicolumn{1}{c}{$P_{orb}$ (hr)} & 
\multicolumn{1}{c}{Spectral Type} &
\multicolumn{1}{c}{$M_1 (M_\odot)$} & 
\multicolumn{1}{c}{$M_2 (R_\odot)$} & 
\multicolumn{1}{c}{$R_2 (R_\odot)$} & 
\multicolumn{1}{c}{References} \\
\hline
V518 Per (GRO J0422+32, Nova Per 1992) & 5.09 	& M2$\pm$2V  & 	&  & 	& 1,2	\\
MM Vel (Nova Vel 1993)	&	6.86	& early K &	&	&	& 3	\\
V616 Mon (A0620-00)	&	7.75	& K4V	& 3.89-4.12 &0.19-0.32 & 0.53-0.63 & 4,5	\\
QZ Vul (GS 2000+25)	&	8.26	& K3-6V & 6.04-13.9 &0.26-0.59	& 0.62-0.81 & 6	\\
GU Mus (Nova Mus 1991)	& 	10.4	& K3-4V	& $6.98\pm1.45$ & $0.94\pm0.40$	& $1.06\pm0.15$	& 7,8,9\\
V2107 Oph (Nova Oph 1977)& 	12.5	& K5V	&  &	& 	& 10	\\
V822 Cen (Cen X-4) 	& 	15.1    & K5V & &	&	& 11,12	\\
V1333 Aql (Aql X-1) 	&	19.0 	& K5V   &  &	&	& 11	\\
V1033 Sco (GRO J1655-40, Nova Sco 1994)& 62.9	& F3-6IV & $7.02\pm0.22$ &$2.34\pm0.12$ & $4.85\pm0.08$ & 13\\
V404 Cyg (GS 2023+338)	& 155.3	& K0IV & $12^{+3}_{-2}$ & $0.7^{+0.3}_{-0.2}$ & $6.0^{+0.7}_{-0.5}$ & 14,15\\
\hline
\end{tabular}
\begin{tabular}{l}
References:
1. \pcite{chevalier96},
2. \pcite{casares95a},
3. \pcite{shahbaz96b},
4. \pcite{mcclintock86},\\
\hspace{1.46cm}
5. \pcite{marsh94},
6. \pcite{harlaftis96},
7. \pcite{orosz96},
8. \pcite{casares97},\\
\hspace{1.46cm}
9. This paper.
10. \pcite{harlaftis97},
11. \pcite{shahbaz96c},
12. \pcite{mcclintock90},\\
\hspace{1.46cm}
13. \pcite{orosz97},
14. \pcite{casares94a},
15. \pcite{shahbaz94b}.\\
\end{tabular}
\end{table*}

%

\section{The mass--radius relation}
\label{sec:mr}

\begin{figure*}
\label{fig:mr}
\psfig{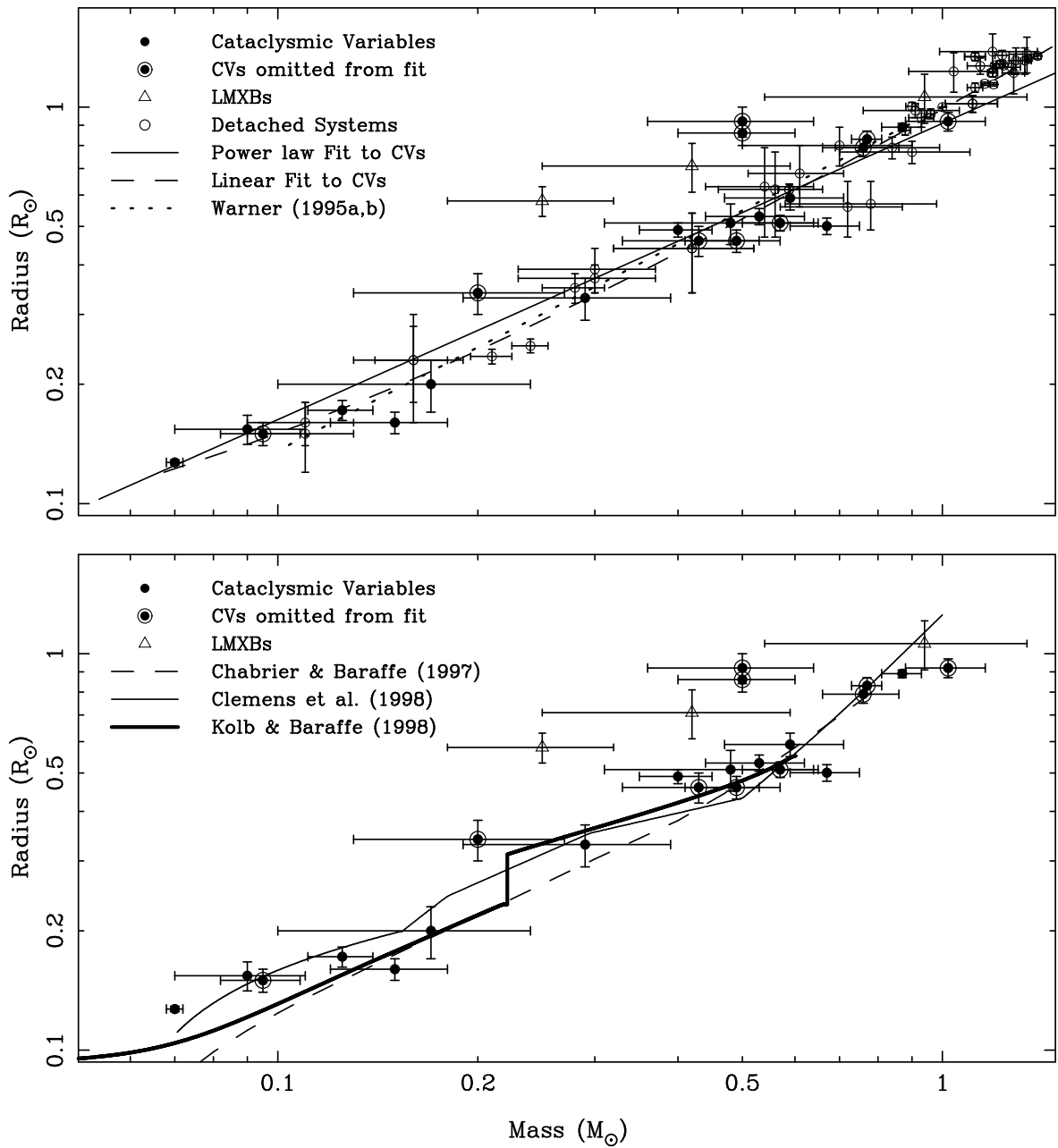}
\caption{The masses and radii of the secondary stars in CVs and LMXBs. The upper panel shows 
a power law fit (solid line) and a linear fit (dashed line) to the CV data (unringed, filled 
circles). The CV points which are ringed have been omitted from the fit. LMXBs are denoted 
by open triangles. Also plotted in the upper panel is the mass--radius relation (equation~7)
derived by Warner (1995a,b). The lower panel shows the same data points as the upper panel, 
along with the theoretical models of Chabrier \& Baraffe (1997, the dotted line) and the 
empirical relation derived by Clemens et al. (1998, the solid line). The thick, solid line 
shows the secular evolution of the mass and radius of the secondary star computed by 
\protect\scite{kolb98a}. 50 stars in detached binaries with well-determined masses and radii 
have been plotted as open circles in the upper panel. See text for details.}
\end{figure*}

The masses of the component stars in CVs can be determined in a number of different ways, 
using measurements of:

\begin{enumerate}
\item 
The radial velocity semi-amplitude of the white dwarf, $K_W$. This is inferred from the 
velocity variations of the wings of the emission lines, which arise in the inner accretion 
regions and are assumed to follow the motion of the white dwarf. These measurements are 
often unreliable because of contamination from other emission-line sources, such as the 
bright spot and the irradiated face of the secondary star, which do not follow the motion 
of the white dwarf and hence introduce phase shifts in the radial velocity curves. It is 
possible to correct for this contamination using diagnostic diagrams \cite{shafter86a}, 
light centre diagrams \cite{marsh88a} and symmetry analyses of Doppler tomograms \cite{still96}.
However we do not accept that any of these methods do give truly reliable values of $K_W$ 
because of the uncertainty in the extrapolation needed to correct for the large phase shifts 
in the radial velocity curves.

\item
The radial velocity semi-amplitude of the secondary star, $K_R$. This is measured using 
the motion of absorption lines such as the Na{\small I} $\lambda8190$\AA\ doublet (e.g. 
Friend et al. 1990a,b) or by using skew mapping \cite{smith98}. Another option is to 
follow the motion of the secondary star using the line emission from its irradiated inner 
face (e.g. \pcite{beuermann90}). All of these $K_R$ measurements are subject to errors 
due to the non-uniform distribution of the emission/absorption line strength on the 
surface of the secondary star, although this can be corrected for (e.g. \pcite{wade88}, 
\pcite{rutten94a}). 

\item
The projected rotational velocity of the secondary star, $V \sin i$. This is usually 
measured through comparison with (slowly rotating) field star template spectra which 
are given a series of artificial rotational broadenings. A constant times each template 
spectrum is then subtracted from the object spectrum to give a residual spectrum. The 
template which yields the smoothest residual provides the value of $V \sin i$ (e.g. 
\pcite{marsh94}, \pcite{smith98}). An alternative method of measuring $V \sin i$ is 
that of cross-correlating with a template spectrum and measuring the width of the 
cross-correlation peak, $\sigma_{rot}$ (e.g. \pcite{horne86b}). Models of rotating 
stars with various amounts of limb darkening are then used to find values of $V\sin i$ 
corresponding to $\sigma_{rot}$. Measurement of $V \sin i$ provides a powerful constraint 
on $M_2$, as the relation between them is only weakly dependent on $q$ (see equation~19 
in Appendix B).

\item
The full width of the eclipse at half depth  $\Delta \phi_{1/2}$. At half depth it is 
assumed that one half of the accretion disc is eclipsed, corresponding to the point at 
which the white dwarf is eclipsed. This ceases to be true if the disc is asymmetrical, 
e.g. if there is a dominant bright spot, or if there is no disc, e.g. in magnetic CVs. 
If the system is non-magnetic and does not have a dominant bright spot then the measurement 
of $\Delta \phi_{1/2}$ provides a relationship between $q$ and the inclination $i$ (e.g. 
\pcite{smith98}). 

\item
The duration and phase of the ingress/egress of the bright spot. This assumes that 
the bright spot lies on a ballistic trajectory from the $L_1$ point, i.e. there is 
no significant magnetic field. This provides another relationship between $q$ and $i$ 
(e.g. \pcite{wood89}).

\item
The duration of the ingress/egress of the white dwarf, $\Delta \phi_{wd}$. This is a 
measure of the radius of the primary as a function of $i$ and the separation $a$. The 
presence of an extended boundary layer can, however, cause the radius of the white dwarf 
to be over-estimated and hence the mass to be under-estimated (e.g. \pcite{wood89}). 

\item
The radius of the orbit of the white dwarf about the centre of mass, $a_{wd}$, for rapidly 
rotating white dwarfs. This can be estimated from the spin pulse delay (e.g. AE~Aqr, 
\pcite{eracleous94}). The only question is whether the white dwarf is the source of the 
pulses; in AE~Aqr the spin pulse delay is exactly in phase with the white dwarf so there 
is little doubt.

\item
The ellipsoidal variations due to the changing aspect of the distorted Roche lobe-filling 
secondary star. These are particularly prominent in the infrared, showing up as a distinctive 
double-humped modulation in the light curve. With appropriate modelling it is possible to 
constrain the inclination using ellipsoidal variations. This technique is especially useful 
for non-eclipsing systems (e.g. \pcite{hilditch95}). The main uncertainty with this technique
is the contribution of the disc to the total flux, which reduces the amplitude of the
variations, resulting in an underestimate of the inclination. This can be corrected for if 
the secondary star has been detected spectroscopically (e.g. \pcite{shahbaz96d}).

\item
Linear polarisation light curves of magnetic CVs. These can be modelled to constrain $i$ 
(e.g. \pcite{wickramasinghe91}).

\item
The width of the base of the emission lines. This gives the projected rotational velocity 
of the innermost parts of the accretion disc and hence a relation between $M_1$, $R_1$ 
and $i$ (e.g. \pcite{kuerster88}). This quantity is, however, difficult to measure with 
any precision.

\item
The separation of the two peaks in the emission lines, $v_D$. This is used to relate the 
accretion disc radius (which must be less than the radius of the primary star's Roche lobe) 
and the mass of the primary star (e.g. \pcite{horne86c}). In combination with $K_R$, this 
can provide an upper limit to $q$.

\item
The orbital modulations of $V \sin i$. With several assumptions about limb and 
gravity darkening these modulations can be modelled to constrain the inclination 
(e.g. \pcite{casares96}, \pcite{shahbaz98a}). This technique is really a subset 
of mass derivations using Roche tomography \cite{rutten96}, which utilise the 
modulations in line strength and position as well as width to determine the 
component masses.

\item
A mass--radius relation for the white dwarf (e.g. \pcite{hamada61}, \pcite{nauenberg72}). 
This is commonly used in conjunction with item (vi).

\item
A main-sequence mass--radius relation for the secondary star. The assumption of a main 
sequence mass--radius relation is made frequently in the literature, and it is the validity
of this assumption which we wish to test.
\end{enumerate}

Of these various measurements and techniques for mass determination, our preferred method is
the combination of $K_R$, $V \sin i$ and $\Delta\phi_{1/2}$, or if the system is non-eclipsing,
ellipsoidal variations. These parameters are simple to measure if the secondary is sufficiently 
bright, do not depend on any assumptions, other than that the secondary fills its Roche lobe,
and can have any biases (due to irradiation of the secondary, for example) corrected for quite 
straightforwardly (e.g. \pcite{davey96}).

\scite{ritter98} list values for the mass of the secondary star in 81 CVs. The vast 
majority of these mass determinations, however, have been derived using a main sequence 
mass--radius relation. As our goal is the derivation of a new mass--radius relation for 
CV secondary stars, we have been forced to omit from our consideration all of the mass 
estimates which make a main sequence assumption. This leaves only 20 mass determinations. 
Of these, 5 use the width of the emission lines (x and xi), which we believe to be extremely 
unreliable due to measurement uncertainties, 2 use the eclipse method (iv,v,vi) when it 
is clear from the shape of the light curve that this is invalid or at least unreliable, 
and 6 use determinations of $K_W$ which are derived from radial velocity curves with 
significant phase shifts and hence do not represent the motion of the white dwarf. In 
fact, the only mass determination wholly dependent on a $K_W$ measurement which we have 
not rejected is that of IP~Peg, as \scite{marsh88a} was able to successfully correct for 
the phase shift in the radial velocity curve using the light centres method.

The above filtering process leaves us with only 7 reliable CV mass determinations listed in 
the \scite{ritter98} catalogue. We have also uncovered a few additional mass determinations 
which are not listed in their catalogue. These are BT~Mon, V895~Cen and V2051~Oph. We have also 
discovered a number of CVs in the literature which possess measurements of $K_R$, $V \sin i$, 
$\Delta\phi_{1/2}$ and $i$ but which have had no mass calculations -- for these objects 
(BD~Pav, DX~And, EX~Dra, AM~Her and the LMXB GU~Mus), we performed Monte Carlo simulations 
similar to those described by \scite{smith98} to calculate the system parameters -- these 
are listed in Appendix~C. 

In summary, we have 14 reliable CV mass determinations and 8 less reliable mass determinations, 
the latter depending on sometimes dubious $K_W$ measurements, or else the uncertain assumption 
that the eclipse method is valid (UU~Aqr). The masses and radii are
listed in Table~3, and described in greater detail in Appendix~A. Note also that a number 
of authors quoted only the masses of the secondary stars and not their radii -- we have 
calculated the radii using equation~(3). Where this is the case, the errors on the radii have 
been taken to be a third of the percentage errors on the mass (from equation~3).

There are just 5 LMXBs listed by Ritter \& Kolb (1998, see also \pcite{beekman97}) which have 
reliable mass determinations. We have determined the system parameters of GU~Mus using values 
of $V\sin i$ and $K_R$ provided by Casares et al. (1997; who only determined upper and lower 
limits to the masses). The masses and radii of the LMXB secondary stars are listed in Table~2 
and described in more detail in Appendix~A. 

\begin{table*}
\label{tab:mr}
\caption{Masses and radii of cataclysmic variables. Key: DN = dwarf nova; CN = classical nova; 
NL = nova-like; P = polar; IP = intermediate polar. Asterisks denote those stars whose system 
parameters were wholly dependent on $K_W$ measurements derived from radial velocity curves 
exhibiting significant phase shifts.}

\begin{tabular}{lccccccl} 
\hline
\multicolumn{1}{l}{Star} &
\multicolumn{1}{c}{Class} & 
\multicolumn{1}{c}{$P_{orb}$} & 
\multicolumn{1}{c}{$M_1$} &
\multicolumn{1}{c}{$q = M_2/M_1$} &
\multicolumn{1}{c}{$M_2$} &
\multicolumn{1}{c}{$R_2$} &
\multicolumn{1}{l}{Refs} \\
\hline
V2051 Oph &	DN 	& 1.50 & $0.78\pm0.06$ & $0.19\pm0.03$	& $0.15\pm0.03$ & $0.16\pm0.01$ &	1 \\
OY Car	  &	DN 	& 1.51 & $0.685\pm0.011$ & $0.102\pm0.003$& $0.070\pm0.002$ & $0.127\pm0.002$ &	2 \\
EX Hya*   & 	IP 	& 1.64 & $0.49\pm0.03$ & $0.19\pm0.03$	& $0.095\pm0.013$ & $0.15\pm0.01$ &  	3, 4, 5 \\
HT Cas	  &	DN 	& 1.77 & $0.61\pm0.04$ & $0.15\pm0.03$	& $0.09\pm0.02$ & $0.154\pm0.013$ &	6 \\	
Z Cha	  &	DN 	& 1.79 & $0.84\pm0.09$ & $0.20\pm0.02$	& $0.125\pm0.014$ & $0.172\pm0.010$ &	7, 8 \\	
ST LMi	  &   	P 	& 1.89 & $0.76\pm0.30$ & $0.23\pm0.05$ 	& $0.17\pm0.07$ & $0.20\pm0.03$ &  	5, 9, 10\\
AM Her    &	P	& 3.09 & $0.44\pm0.12$ & $0.64\pm0.10$ 	& $0.29\pm0.10$ & $0.33\pm0.04$ &  	5, 11, 12, 13\\
IP Peg 	  &	DN 	& 3.80 & $1.15\pm0.10$ & $0.59\pm0.04$	& $0.67\pm0.08$ & $0.501\pm0.024$ &	14, 15 \\	
CN Ori*   &	DN 	& 3.92 & $0.74\pm0.1$ &  $0.66\pm0.04$	& $0.49\pm0.08$ & $0.46\pm0.03$ &	16 \\	
UU Aqr*	  &	NL 	& 3.93 & $0.67\pm0.14$ & $0.33\pm0.10$	& $0.20\pm0.07$ & $0.34\pm0.04$ &	17 \\	
U Gem*    &	DN 	& 4.25 & $1.26\pm0.12$ & $0.46\pm0.03$	& $0.57\pm0.07$ & $0.510\pm0.023$ &	16, 18 	\\	
BD Pav*	  &	DN 	& 4.30 & $0.95\pm0.10$ & $0.44\pm0.06$	& $0.43\pm0.10$ & $0.46\pm0.04$ &  	5, 16, 19\\
DQ Her	  &	CN,IP	& 4.65 & $0.60\pm0.07$ & $0.66\pm0.04$ 	& $0.40\pm0.05$ & $0.49\pm0.02$ &	20, 21 \\	
IX Vel	  &    	NL	& 4.65 & $0.80^{+0.16}_{-0.11}$ &$0.65\pm0.04$ & $0.52^{+0.10}_{-0.07}$ & $0.530\pm0.025$ & 22 \\
V895 Cen  & 	P	& 4.77 & $0.93\pm0.17$ & $0.51\pm0.12$	& $0.48\pm0.17$ & $0.51\pm0.06$	&	23 \\	
EX Dra    &  	DN	& 5.04 & $0.70\pm0.10$ & $0.84\pm0.12$	& $0.59\pm0.12$ & $0.59\pm0.04$ &  	5, 24, 25\\
EM Cyg*   &	DN	& 6.98 & $0.56\pm0.05$ & $1.35\pm0.16$	& $0.76\pm0.10$ & $0.79\pm0.04$ &  	5, 26\\
AC Cnc*   &	NL	& 7.21 & $0.82\pm0.13$ & $1.24\pm0.08$	& $1.02\pm0.14$ & $0.92\pm0.05$ &	27, 28 	\\
V363 Aur* & 	NL	& 7.71 & $0.86\pm0.08$ & $0.89\pm0.03$	& $0.77\pm0.04$ & $0.83\pm0.04$ &	29 \\	
BT Mon	  &	CN	& 8.01 & $1.04\pm0.06$ & $0.84\pm0.04$	& $0.87\pm0.06$ & $0.89\pm0.02$ &	30 \\	
AE Aqr	  &	IP  	& 9.88 & $0.79\pm0.16$ & $0.630\pm0.012$& $0.50\pm0.10$ & $0.86\pm0.06$ &  	5, 31, 32\\
DX And	  &	DN	&10.60 & $0.51\pm0.12$ & $0.98\pm0.10$	& $0.50\pm0.14$ & $0.92\pm0.08$ &  	5, 33, 34\\
\hline
\end{tabular}

\begin{tabular}{l}
References:
1. \pcite{baptista98},
2. \pcite{wood89},
3. \pcite{hellier96b},
4. \pcite{sterken83},
5. This paper,\\
\hspace{1.465cm}
6. \pcite{horne91b},
7. \pcite{wade88},
8. \pcite{robinson95},
9. \pcite{shahbaz96a},\\
\hspace{1.32cm}
10. \pcite{cropper86},
11. \pcite{davey96},
12. \pcite{southwell95},
13. \pcite{wickramasinghe91},\\
\hspace{1.32cm}
14. \pcite{martin89},
15. \pcite{barrera89},
16. \pcite{friend90a},
17. \pcite{baptista94},\\
\hspace{1.32cm}
18. \pcite{smak93},
19. \pcite{harrop96},
20. \pcite{zhang95},
21. \pcite{horne93},\\
\hspace{1.32cm}
22. \pcite{beuermann90},
23. \pcite{buckley98},
24. \pcite{billington96},\\
\hspace{1.32cm}
25. \pcite{fiedler97},
26. \pcite{stover81a},
27. \pcite{okazaki82},\\
\hspace{1.32cm}
28. \pcite{schlegel84},
29. \pcite{schlegel86},\\
\hspace{1.32cm}
30. \pcite{smith98},
31. \pcite{casares96},
32. \pcite{welsh93},\\
\hspace{1.32cm}
33. \pcite{drew93},
34. \pcite{hilditch95}.\\
\hspace{1.32cm}

\end{tabular}
\end{table*}
\rm

Those CVs with reliable mass determinations have been plotted as filled circles in Fig.~2, 
while those which use dubious $K_W$ determinations are ringed. The 
points representing DX~And and AE~Aqr are also ringed as they have evolved secondary 
stars (\pcite{drew93}, \pcite{casares96}). LMXBs are plotted as open triangles. Open 
circles represent the masses and radii of the 50 isolated stars listed by \scite{andersen91} 
and \scite{popper80}, the former having smaller error bars than the latter. 

We performed least-squares power law and linear fits to the CVs using the formal error 
bars quoted in the literature and listed in Table~3. We have only fitted to the unringed, 
filled circles in Fig.~2, i.e. DX~And, AE~Aqr and those systems which have dubious $K_W$ 
measurements have been omitted from the fits. The power law fit is plotted as a solid line 
in the upper panel, and is given by
\begin{equation}
{R \over R_\odot} = (0.91\pm0.09)\Bigl({M \over M_\odot}\Bigr)^{(0.75\pm0.04)}.
\end{equation}
The linear fit is plotted as a dashed line in the upper panel, and is given by
\begin{equation}
{R \over R_\odot} = (0.93\pm0.09)\Bigl({M \over M_\odot}\Bigr)+(0.06\pm0.03).
\end{equation}
We have also plotted the mass--radius relation of Warner (1995a,b), 
\begin{equation} 
R=M^{13/15},
\end{equation}
in the upper panel of Fig.~2 
as a dotted line, which is an approximate fit to the data set given by \scite{webbink90}. 
Of these three mass-radius relations (equations 5,6,7), we would recommend the use of our 
equation 5, which has been fitted to our data using a least-squares fitting procedure
(as opposed to equation 7, where Warner fitted the Webbink's data by eye, forcing 
the multiplicative constant to be unity and the power to be a simple ratio).

In the lower panel of Fig.~2, the dotted line is the theoretical lower main sequence of 
\scite{chabrier97}, calculated using detailed models with the latest input physics, and the 
thin, solid line is the empirical mass--radius relation obtained from a volume-limited sample 
of nearby M-dwarfs by Clemens et al. (1998, see also \pcite{reid97}). The thick, solid line 
shows the masses and radii of a secondary star in an evolutionary sequence computed by 
\scite{kolb98a}, again using the most up-to-date stellar input physics for low-mass stars 
and brown dwarfs currently available. This particular sequence uses a white dwarf mass of 
$M_1 = 0.7 M_\odot$ and an initial secondary star mass of $M_2 = 0.6 M_\odot$. The main 
features in the upper part of the curve are the large radii at high masses (compared to 
the theoretical main sequence) due to thermal inequilibrium and the detachment of the 
secondary from its Roche lobe at $M_2 = 0.22 M_\odot$, which is the point at which (in 
this model) the secondary becomes fully convective and magnetic braking is assumed to 
cease. During this detached phase, there is a 25 per cent fall in radius as the secondary 
star relaxes back to thermal equilibrium. 

The lower part of the curve follows the theoretical main sequence, with the secondary star 
close to thermal equilibrium at low mass transfer rates, until the minimum mass for nuclear 
fusion is reached and the secondary becomes a degenerate, brown dwarf-like object with 
$R_2 \propto M_2^{-1/3}$. Note that the evolutionary sequence should be used with caution 
when interpreting the observed CV masses, as it 
represents the change in mass and radius of a single secondary star with time, whereas the 
observed data is a snapshot of the masses and radii of the population of secondary stars at 
this moment in time. To make a proper comparison with theory, one would need to compute the 
evolutionary tracks of a sample of secondary stars generated by a population synthesis code. 

As a group, the CVs appear to lie slightly above the theoretical main sequence of \scite{chabrier97} 
but appear to fit the main sequence as defined by the distribution of detached systems very well. 
There are a few departures; as already noted, DX~And and AE~Aqr have evolved secondary stars and 
lie above the main sequence, while UU~Aqr and IP~Peg both lie above the main sequence, being 
undersized for their masses, but within the error bars they are still consistent with the main 
sequence. Rather surprisingly, those systems which we have omitted from the fitting process 
because of dubious $K_W$ measurements also follow the main sequence very closely. Unfortunately, 
the errors on the masses and the scatter of the points in Fig.~2 do not allow us to say if the 
secondary stars in CVs follow the disrupted magnetic braking model, or the kinked lower main 
sequence of \scite{clemens98}.\footnote{This has recently been shown not to create a period gap 
at all, but rather two spikes at $P \sim2$\,hr and $P \sim 3$\,hr with the probability of discovering 
a CV in the gap no lower than discovering one outside the gap \cite{kolb98b}.}This means our data are 
unable to test which of these two period gap formation mechanisms is correct (see Section~7). There 
is also no evidence in these data for systems which have evolved beyond the orbital period minimum 
and have low-mass, degenerate brown dwarf-like secondaries (see \pcite{howell97}).

Of the three short period LMXBs, V616~Mon and QZ~Vul both contain secondary stars which lie above 
the main sequence and are somewhat evolved, while GU~Mus lies just on the main sequence (to within 
the error). The secondary stars in the long period LMXBs V404~Cyg and V1033~Sco are both evolved 
(V1033~Sco is also very massive) and do not appear in Fig.~2.
 

\section{The mass--spectral type relation}

An assumption often made in the estimation of system parameters in CVs is that the spectral type of 
the secondary star can be used to estimate its mass. In Fig.~3 we have
plotted those CVs and LMXBs which have measured masses and spectral types given in Tables~1, 2 and 3.  
Also plotted are the detached systems from \scite{andersen91} and \scite{popper80}.

The mass of the secondary is thought to be the most important factor in determining its effective 
temperature, and therefore spectral type \cite{king98}. \scite{stehle96} show that simple homology
relations lead to a relationship between the effective temperature of the star (and therefore
spectral type) and the radius of the form $T_{eff} \propto R^\mu$, where $|\mu| <\!< 1$. Stars should 
therefore expand or contract without changing their spectral types by much and the spectral type
of a star should be almost completely dependent on its mass. 

Fig.~3 shows that this is not the case, and that there is a huge range in mass for 
a given spectral type. Around M5, for example, lie IP~Peg (M4.5) and Z~Cha (M5.5); the secondary star
in IP~Peg is five times more massive than that in Z~Cha. Among the LMXBs there is also little or no 
correlation between spectral type and masses; GU~Mus (K3-4V) is around four times more massive than 
V616~Mon (K4V). The large scatter in masses for a given spectral type are, however, shared by the 
detached systems (e.g. around M4-5 and K0). So although the secondary stars in CVs and LMXBs are not 
too different from main sequence stars, there is too much variation to use the spectral types of CV 
and LMXB secondaries to estimate the secondary star masses, as is also the case with the detached
stars. 

\begin{figure}
\label{fig:mass_sp}
\psfig{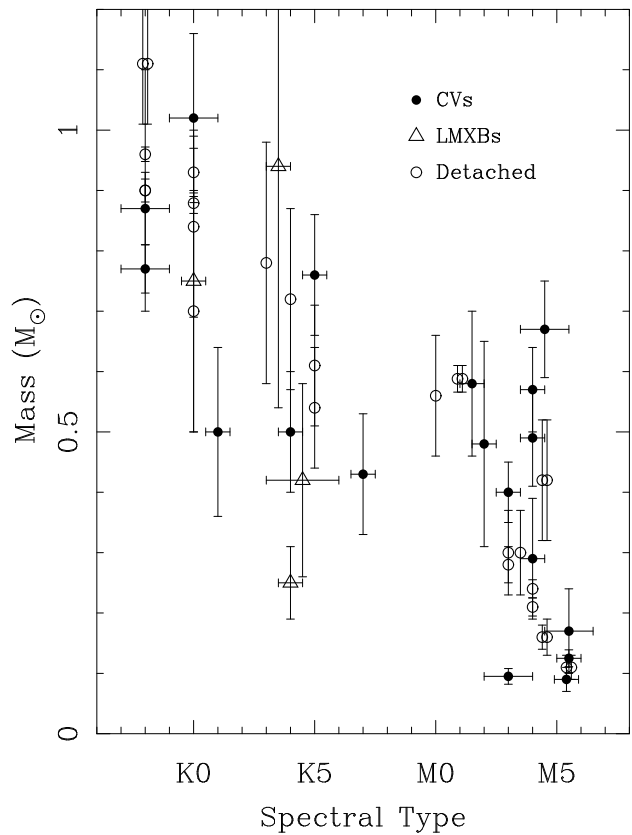}
\caption{The masses and spectral types of the secondary stars in CVs and LMXBs. The filled circles 
represent the CVs, the triangles the LMXBs, and the unfilled circles the detached systems.} 
\end{figure}
 

\section{The mass--period and radius--period relations}

\begin{figure*}
\label{fig:massperiod}
\psfig{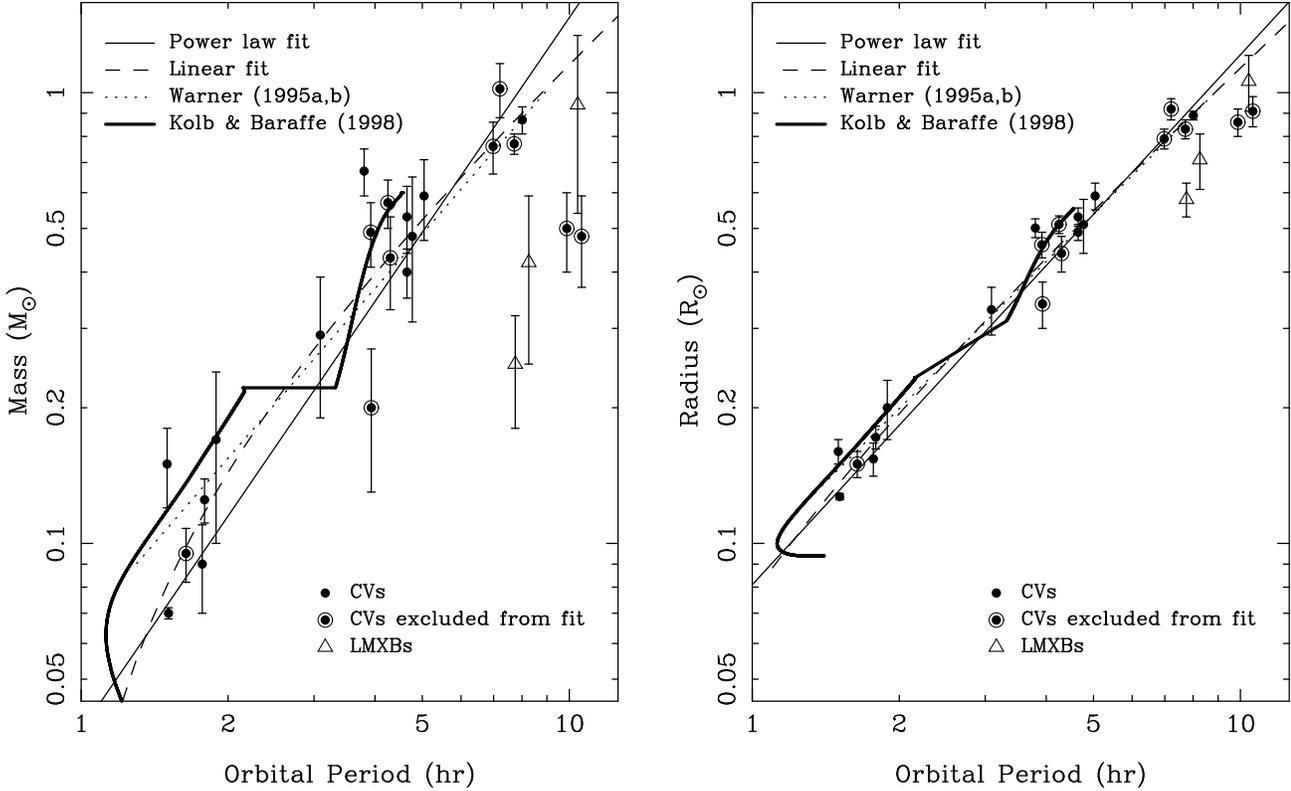}
\caption{The masses and radii of the secondary stars in CVs versus their orbital periods. In each 
panel, the solid line is the power law fit to the CV data (equations 8 and 11) and the dashed 
line is a linear fit to the CV data (equations 9 and 12). The dotted lines are the semi-empirical 
relations of Warner (1995a,b; equations 10 and 13). The thick solid line is the evolutionary sequence 
of \protect\scite{kolb98a}. The ringed points have been omitted from the fits -- see text for details.}
\end{figure*}
 
Another relation of interest is that between the orbital period and the mass of the secondary star. 
We have used the masses and periods listed in Table~3 to derive a mass--period relation. By fitting 
a power law to the data (excluding the evolved secondaries in DX~And and AE~Aqr and the systems with 
dubious $K_W$ determinations), we have derived the following relationship:
\begin{equation}
{M \over M_\odot} = (0.038\pm0.003) P^{(1.58\pm0.09)}.
\end{equation}
We have also performed a linear fit and derived the following mass--period relationship,
\begin{equation}
{M \over M_\odot} = (0.126\pm0.011)P - (0.11\pm0.04).
\end{equation}
Warner (1995a,b) derived the following semi-empirical mass--period relation, which utilised 
his mass--radius relation (equation~7):
\begin{equation}
{M \over M_\odot} = 0.065 P^{5/4}.
\end{equation}

The mass--period relation is shown in the left-hand panel of Fig.~4. The filled circles 
represent the CVs; those which are ringed have been omitted from the fit either because
they depend upon dubious $K_W$ measurements, or because they are evolved to some degree.
The solid line is the power law fit (equation~8), the dashed line is the linear fit 
(equation~9), and the dotted line is Warner's relation (equation~10). The 
evolutionary sequence of \scite{kolb98a} is plotted as a thick, solid line. 

The linear fit (unweighted rms deviation = 0.10) is surprisingly better than the power law fit
(rms = 0.12); one would have expected a homologous relation between mass and period, 
and therefore a power law to be the best fit. Warner's fit is also surprisingly good 
(rms = 0.11), but it is a poorer fit to the systems with short periods and those with 
the most accurate mass determinations. The evolved systems DX~And and AE~Aqr lie a long 
way from the fit, while IP~Peg is twice as massive as the fits predict. The LMXBs all 
have lower masses than the CV fits predict. The period gap is represented in Fig.~4 by 
the flat section of the evolutionary sequence at $M = 0.22M_\odot$; the data are 
insufficiently accurate or numerous to say whether the secondary stars follow it. There 
are no stars lying on the degenerate secondary star arm of the evolutionary sequence 
implying that none of the short period systems in our sample containing the degenerate,
brown dwarf-like secondaries we expect in post-period minimum CVs. 

A similar procedure to that above has been followed to derive a radius--period relation for CV secondary 
stars. A power law fit has been applied to the data in Table~3, again excluding DX~And and AE~Aqr and 
the systems with dubious $K_W$ measurements. The resulting radius--period relation is
\begin{equation}
{R \over R_\odot} = (0.081\pm0.019) P^{(1.18\pm0.04)}.
\end{equation}
We have also performed a linear fit, which is given by
\begin{equation}
{R \over R_\odot} = (0.117\pm0.004) P - (0.041\pm0.018).
\end{equation}
The corresponding semi-empirical radius--period relation of Warner(1995a,b) is
\begin{equation}
{R \over R_\odot} = 0.094 P^{13/12}.
\end{equation}

The right hand panel in Fig.~4 shows the radius--period relation for CVs. The solid line is the 
power law fit (equation~11) to the CV data (unringed circles), the dashed line is the linear fit 
(equation~12) and the dotted line is Warner's relation (equation~13). The thick solid line is 
the evolutionary sequence of \scite{kolb98a}. The rms scatters from the fits show them to be of 
similar quality; 0.033 for the linear fit, 0.042 for the power law and 0.036 for Warner's relation. 
The correlation here is stronger than that in the mass--period diagram, 
as is to be expected: from equation~(3), any deviation in mass from the mass--period relation 
must be matched by a deviation one third the size in the radius--period diagram.
The LMXBs have smaller radii than their orbital periods predict, again the deviation from the fit 
is around one third of that in the mass--period plot.


\section{White dwarf masses}

\begin{figure}
\label{fig:wdper}
\psfig{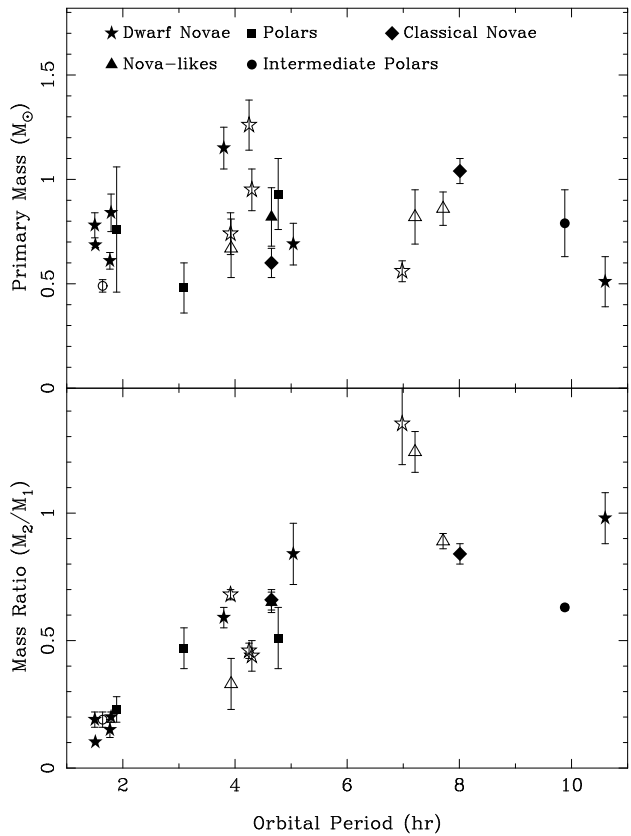}
\caption{The masses of the white dwarfs and the mass ratios of CVs plotted against orbital period. 
Open symbols are used for those mass determinations which used unreliable $K_W$ measurements.}
\end{figure}

The careful selection of systems with well measured secondary star masses and radii also provides 
us with the most accurately determined sample of CV white dwarf masses. In Table~4, we list the 
mean white dwarf mass for each of the CV subtypes, above and below the period gap, along with the 
standard errors, followed by the number of systems in parentheses. We have computed weighted and 
unweighted means for the white dwarfs -- the methods used are given in the footnotes to Table~4. 
The unweighted mean is probably the more reliable statistic to use, since the range of formal 
errors in the white dwarf mass determinations is so great; the photometric determinations which 
have small formal errors but certainly much larger systematic errors (e.g. OY~Car) dominate in 
the calculation of the weighted means. The dispersions on the unweighted means are also more 
representative than those quoted for the weighted means. White dwarf masses are plotted against 
orbital period in the upper panel of Fig.~5, with each CV subtype represented by a different symbol. 

The systems below the period gap have a lower mean mass than those above, which is to be 
expected, since high mass white dwarfs are required to support stable conservative mass 
transfer from the more massive secondaries which reside in long period CVs. Most of the 
white dwarf masses are consistent with CO white dwarfs; the white dwarfs in U~Gem and 
IP~Peg are near the minimum mass for ONeMg white dwarfs ($\sim 1.15M_\odot$, e.g. 
\pcite{iben97}), but none are light enough to be He white dwarfs ($\lta 0.4M_\odot$). The 
mean mass for systems below the gap is probably more representative of the class as a whole, 
since strong selection effects come into play: it is thought that up to 99 per cent of the 
actual CV population are below the gap (\pcite{kolb93} - but see also \pcite{patterson98}), 
despite the fact that they comprise only about 40 per cent of those observed. The white 
dwarf masses show no trends with CV sub-type.

\scite{webbink90} performed statistics on a larger sample of white dwarf masses, but a large 
number of these determinations were unreliable, depending on measurements of emission 
line profiles (FWHM, double-peak separation, rms line-widths) which had been calibrated against 
a few well observed double-lined CVs and Algols. It is, however, worth comparing Webbink's figures 
with ours as his mean white dwarf masses are the most often quoted. The mean masses derived by 
\scite{webbink90} for all systems, and for those above and below the period gap, are listed in 
Table~4. We reach the same general conclusions as Webbink, the mean white dwarf mass above the 
period gap is higher than that below, and that the overall mean mass is $0.76\pm 0.22 M_\odot$, 
higher than the mean mass of single white dwarfs ($0.6 M_\odot$, e.g. \pcite{bergeron95}).

The mass ratio--orbital period diagram in the lower panel of Fig.~5 shows the near linear relation 
we expect, since the secondary stars' masses are well correlated with their orbital periods. The scatter 
in this is largely due to the scatter in white dwarf masses. Again there is no trend according to 
subtype; the two systems with anomalously large mass ratios, EM~Cyg and AC~Cnc, are a dwarf nova 
and a novalike respectively. Note that this sample of accurate mass ratio determinations is not 
exhaustive, there exist several systems which have the mass ratio determined, but which do not 
have accurate mass determinations (e.g. because the inclination is unknown.) 

In our sample, the unweighted mean value of the mass ratio (performed on those systems which do not 
have dubious $K_W$ values) is $\bar q = 0.17\pm0.05$ below the period gap and $\bar q = 0.70\pm0.15$ 
above the period gap. These values are consistent with the statistical analysis of eclipse durations 
by \scite{bailey90}, who found that the mean value of the mass ratio should be $\bar q = 0.13\pm0.03$ 
below the gap and $\bar q = 0.65\pm0.12$ above the gap. 

\begin{table*}
\label{tab:wd2}
\caption{The mean white dwarf masses for each of the CV subtypes, for systems 
below the period gap, above the period gap, and in total. The standard errors are also listed, 
with the numbers of systems used in determining the mean given in parentheses. Note that DQ~Her 
has been counted as both a classical nova and an intermediate polar.}

\begin{tabular}{llll}
\hline
\multicolumn{4}{c}{A. Unweighted Average$^a$}\\
\hline
\hline
\multicolumn{1}{l}{Sample} &
\multicolumn{1}{c}{Below period gap} & 
\multicolumn{1}{c}{Above period gap} & 
\multicolumn{1}{c}{All periods} \\
\hline
Dwarf novae	& \,\,$0.73\pm0.10$\,\,(4)	& \,\,$0.84\pm0.29$\,\,(7) & \,\,$0.80\pm0.24$\,\,(11)\\
Nova-likes	& $\;\;\;\;\; -$	& \,\,$0.79\pm0.08$\,\,(4) 	   & \,\,$0.79\pm0.08$\,\,(4)\\
Polars		& \,\,$0.76^*\;\;\;\;\;\;\;\;\;\;$(1) & \,\,$0.69\pm0.35$\,\,(2)& \,\,$0.71\pm0.25$\,\,(3)\\
Intermediate polars & \,\,$0.49^*\;\;\;\;\;\;\;\;\;\;$(1) & \,\,$0.70\pm0.13$\,\,(2) & \,\,$0.63\pm0.15$\,\,(3)\\
Classical novae	& $\;\;\;\;\; -$	& \,\,$0.82\pm0.31$\,\,(2) 		& \,\,$0.82\pm0.31$\,\,(2)\\
All systems	& \,\,$0.69\pm0.13$\,\,(6)	& \,\,$0.80\pm0.22$\,\,(16)	& \,\,$0.77\pm0.21$\,\,(22)\\
\hline
\multicolumn{4}{c}{B. Weighted Average$^b$}\\
\hline
\hline
\multicolumn{1}{l}{Sample} &
\multicolumn{1}{c}{Below period gap} & 
\multicolumn{1}{c}{Above period gap} & 
\multicolumn{1}{c}{All periods} \\
\hline
Dwarf novae	& \,\,$0.68\pm0.01$\,\,(4)	& \,\,$0.75\pm0.03$\,\,(7) & \,\,$0.69\pm0.01$\,\,(11)\\
Nova-likes	& $\;\;\;\;\; -$		& \,\,$0.82\pm0.06$\,\,(4) & \,\,$0.82\pm0.06$\,\,(4)\\
Polars		& \,\,$0.76\pm0.30$\,\,(1)	& \,\,$0.58\pm0.09$\,\,(2) & \,\,$0.60\pm0.09$\,\,(3)\\
Intermediate polars & \,\,$0.49\pm0.03$\,\,(1) 	& \,\,$0.63\pm0.06$\,\,(2) & \,\,$0.52\pm0.03$\,\,(3)\\
Classical novae	& $\;\;\;\;\; -$		& \,\,$0.85\pm0.05$\,\,(2) & \,\,$0.85\pm0.05$\,\,(2)\\
All systems	& \,\,$0.66\pm0.01$\,\,(6)	& \,\,$0.78\pm0.02$\,\,(16)& \,\,$0.68\pm0.01$\,\,(22)\\
\hline
\hline
Webbink (1990) 	& \,\,$0.66\pm0.01$\,\,(26)	& \,\,$0.81\pm0.04$\,\,(58)& \,\,$0.74\pm0.04$\,\,(84)\\
All systems 	& & & \\
\hline
\multicolumn{4}{l}{*No associated error.}\\
\end{tabular}
\begin{tabular}{lllllll}
$^a\;{\overline M}$ & = &${1 \over N} {\sum}^{N}_{i=1} M_i$ &$ \sigma^2_M$ & = &$ {1 \over {N-1}} {\sum^{N}_{i=1}} (M_i - {\overline M})^2$ \\
$^b\;{\overline M}$&=&${\sum^{N}_{i=1}} {{M_i}\over{\sigma^2_i}} / {\sum^{N}_{i=1}} {1\over{\sigma^2_i}}$ & $\sigma^2_M$& = &$ \Bigl({\sum^{N}_{i=1}} {1\over{\sigma^2_i}}\Bigr)^{-1}$ \\
\end{tabular}
\end{table*}

%

\section{Discussion}
\label{sec:discuss}
\subsection{Are CV secondaries main sequence stars?}
The spectral type--period diagram (Fig.~1) shows that the secondary stars in CVs lie very close 
to the main sequence defined by the detached systems. There is little differerence between the 
distribution of CV secondaries and that of the detached stars, up to an orbital period of about 
7--8 hours. Beyond that, for the reasons stated in section~1, the secondary stars are required 
to be evolved, and we see AE~Aqr ($P = 9.88$\,hr) and DX~And ($P = 10.60$\,hr) lying well above 
the rest of the distribution with later spectral type than the fit predicts.

The mass--radius diagram (Fig.~2) clearly shows that the secondary stars in CVs are on the 
whole indistinguishable from the observed main sequence stars in detached binary systems. 
Again, the only outlying points are AE~Aqr and DX~And, which almost certainly have evolved 
secondary stars. The most interesting question from a mass-determination point of view, however, 
is not whether secondary stars as a whole are main sequence stars, but rather at what point can 
we no longer apply a main sequence mass--radius relationship. From Figs.~1~\&~2 we can say that 
CVs with periods of up to 7--8 hours almost always have main sequence secondary stars, although 
one must always beware of the existence of systems with peculiar secondaries and use the 
mass--radius relation with caution (e.g. by correctly propagating the errors in our mass--radius
relation when determining masses).

We have seen that the expected divergence from the main sequence does occur in long periods 
CVs, but there is as yet no firm evidence for post-period minimum CVs with degenerate brown 
dwarf-like secondaries in short period CVs. The evolutionary sequence of \scite{kolb98a} shows 
us where to look, but as yet there have been no brown dwarf-like secondaries detected in CVs 
either by their spectral types (e.g. by looking for spectral types much later than predicted by 
equation 4) or by their position on the mass--radius diagram (but see \pcite{howell98}). This 
does not mean, of course, that there are no post-period minimum CVs; these systems should be 
intrinsically very faint, and near the period minimum (where WZ Sge and many other suspected 
post-period minimum CVs lie) their secondary stars will be difficult to distinguish from
late-type main sequence stars.


\subsection{Are LMXB secondaries main sequence stars?}
The spectral types of the LMXBs seem to follow the same distribution as the CVs with orbital 
period, while their radii are larger than main sequence stars of the same type, suggesting they 
are either somewhat evolved (as predicted by \pcite{king97}), or else have expanded towards a 
new state of thermal equilibrium due to the level of X-ray irradiation to which they are exposed 
\cite{podsiadlowski91}. The upper orbital period limit above which secondary stars cannot be on 
the empirical main sequence rises (from around 9~hours for CVs), to around 15~hours, using the 
theoretical upper neutron star mass limit of $\sim 3 M_\odot$ with a typical A0 main sequence 
star of mass $2.40M_\odot$ and radius $1.87M_\odot$ \cite{gray92}. In LMXB black hole candidates, 
of course, there is no such limit to the mass of the black hole and therefore no limit on the 
secondary star.


\subsection{Consequences for the disrupted magnetic braking model}
The generally accepted mechanism for the formation of the period gap in CVs is the disrupted 
magnetic braking model. Above the period gap magnetic braking is the dominant mechanism for 
the angular momentum loss which drives mass transfer. As the CV moves towards a period of 
3~hours, the secondary star's mass falls to around $0.25M_\odot$, at which point it becomes 
(almost) fully convective \cite{kolb93}. Magnetic braking is then thought to cease, allowing 
the secondary, which due to its high mass transfer rate has been out of thermal equilibrium, 
to relax and shrink inside its Roche lobe. This cuts off mass transfer (and the cataclysmic 
behaviour) until further loss of angular momentum via gravitational radiation reduces the 
period to around 2 hours, at which point the secondary comes back into contact with its Roche 
lobe and recommences mass transfer (\pcite{spruit83}; \pcite{rappaport83}).

A mass--radius diagram would be expected to show a ``kink'' at around a mass of $0.25M_\odot$, as 
the CV secondary stars revert from being out of thermal equilibrium and therefore undermassive 
for their radii above the period gap, to being thermally relaxed at low masses below the period 
gap. This is shown in the secular evolution model plotted in the lower panel of Fig.~2. If a 
star enters the period gap at $P_{above}=3.0$\,hr and emerges at $P_{below}=2.0$\,hr, without 
having lost any further mass, then, using equation~(3), its radius must have shrunk by a factor
\begin{equation}
\Bigl({R_{below} \over R_{above}}\Bigr) = \Bigl({P_{below} \over P_{above}}\Bigr)^{2/3} = 0.763
\end{equation}
i.e. a decrease of almost 25 per cent. A narrower period gap, e.g. 2.2--2.8 hrs, 
would see a smaller drop in the radius at the top of the period gap, in this case 
15~per cent.

We are unable to see the expected sudden drop in Fig.~2, although the data is sparse 
in the crucial region around $0.25M_\odot$, the only system there being AM~Her. In 
this interesting region, just above the period gap, 
there is a dearth of dwarf novae, instead the CV population is dominated by peculiar 
nova-likes such as the SW~Sex stars \cite{dhillon96}, which have intrinsically bright 
discs and almost undetectable secondary stars, and for which determination of system 
parameters is therefore very difficult. 

\scite{clemens98} see a dip in the colour-magnitude diagram for M-dwarfs at $M_V\sim 12$
(around $0.25M_\odot$). They suggest that this means that the radius of the secondary 
star in a CV would shrink more rapidly in this region for the same rate of mass loss, 
leading to the rapid crossing of the 2--3\,hr orbital period range. This would then be 
a possible mechanism for the formation of the period gap. This interpretation has been 
disputed by \scite{kolb98b}, who show that if the mass--radius relation of Clemens et al. 
held, then rather than producing a period gap, the period distribution would have two 
spikes at the upper and lower edges of the ``gap'', with the probability of discovering CVs
inside the ``gap'' the same as discovering them outside the ``gap.''

Unfortunately we are not yet at the stage where we can confirm or disprove the disrupted 
magnetic braking model observationally. Our dataset is simply too sparse and insufficiently 
accurate. Several more accurate measurements of secondary star masses and radii are 
needed, especially around the period gap (in which there are sadly few secondary star 
detections) in order to differentiate between these competing period-gap formation 
theories. 

%

\section{Conclusions}
\begin{enumerate}
\item
We find there are a total of 55 reliable spectral type determinations and only 14 reliable mass determinations
of CV secondary stars (10 and 5, respectively, in the case of LMXBs).
\item 
We have derived new mass--radius, mass--period, radius--period and spectral type--period relations 
for CV secondary stars, using a carefully selected sample of CVs with well measured system parameters.
\item 
The secondary stars in CVs with periods below 7--8 hours are, as a group, indistinguishable from main 
sequence stars in detached systems in terms of spectral type, mass and radius.  
\item 
The secondary stars in LMXBs show some evidence for evolution, with radii which are slightly too large 
for their masses.
\item 
We have shown that the assumption that the spectral type of the secondary star in CVs and LMXBs 
provides a good estimate of its mass is not a good one. 
\item 
We have calculated the mean white dwarf mass in CVs, for the various CV subtypes, both above 
the period gap (where we find ${\overline M}_1 = 0.80\pm0.22M_\odot$) and below the period gap 
(where we find ${\overline M}_1 = 0.69\pm0.13M_\odot$). 
\item 
We have shown that accurate measurements of masses and radii in CV secondary stars can be used 
to constrain CV evolution and provide evidence for or against the disrupted magnetic braking theory.
\item 
We have demonstrated the need for many more accurate CV mass measurements, especially around the 
period gap, to test the disrupted magnetic braking model, and among the faint short period CVs 
to search for the predicted post-period minimum systems.
\end{enumerate}

%

\section*{\sc Acknowledgements}

We are indebted to Mike Politano and Uli Kolb for providing us with secular evolution 
models and Ron Webbink for supplying his dataset of CV system parameters. We would also 
like to thank Uli Kolb and Mike Politano for invaluable discussions on secondary stars 
and the evolution of CVs, and the referee, Hans Ritter, for his careful reading of the 
manuscript and suggestions for improvements. DAS is supported by a PPARC studentship.

\bibliographystyle{mnras}
\bibliography{abbrev,refs}

\section*{Appendix A}
\subsection*{Notes on individual systems}

\begin{enumerate}
\item {\it V2051~Oph. }
\scite{baptista98} used a purely photometric method to find the masses of the components in this 
dwarf nova. The white dwarf and bright spot eclipses are both clearly visible in the light 
curve so the additional measurement of $\Delta\phi_{1/2}$ allowed the system parameters to 
be determined using the white dwarf mass--radius relation of \scite{hamada61} and a Monte 
Carlo simulation. 

\item {\it OY Car. }
\scite{wood89} also used the photometric method to determine the masses in this dwarf nova.
Several other mass determinations cited by \scite{wood89} all give higher values for $M_2$.

\item {\it EX Hya. }
This intermediate polar has a detectable secondary star. \scite{hellier96b} combined $K_W$ 
\cite{hellier87} and $K_R$ (measured using skew mapping; \pcite{smith93b}) with $\Delta\phi_{1/2}$ 
to obtain the mass. We have calculated the radius of the secondary star using equation~(3).

\item {\it HT Cas. }
\scite{horne91b} performed a similar analysis to that of \scite{wood89}.

\item {\it Z Cha. }
\scite{wood86} used the same technique as that applied to V2051~Oph, and obtained values
of $0.081\pm0.003M_\odot$ and $0.149\pm0.002R_\odot$ for the mass and radius of the 
secondary star respectively. 
\scite{wade88} derived another mass estimate, using the mass ratio and inclination $i$ derived 
by \scite{wood86}, but rather than using the white dwarf ingress/egress timings, they obtained 
a measurement of $K_R$. The higher values derived by \scite{wade88} are those used throughout 
this paper. The inconsistency between the two techniques cannot be satisfactorily explained 
by errors in the Hamada-Salpeter relation or the presence of a thick boundary layer above the 
surface of the white dwarf.

\item {\it ST LMi. }
\scite{shahbaz96a} measured the secondary star's radial velocity in this non-eclipsing polar, 
and determined the masses using the value of $i$ derived from polarimetric measurements 
\cite{cropper86}.

\item {\it AM Her. }
We have calculated the mass and radius of the secondary in the prototypical polar, 
using the values of $K_R$ and $V \sin i$ determined by \scite{southwell95}, with the 
appropriate K-correction given by \scite{davey96}, and $i$ determined using polarimetry 
by \scite{wickramasinghe91}. \scite{southwell95} found an 8 per cent bias in their 
measurement of $V\sin i$, we therefore adopt $V \sin i = 100 \pm 10$\,km\,s$^{-1}$. 
Note that \scite{shahbaz96a} found $V \sin i = 68 \pm 12$\,km\,s$^{-1}$, using fewer 
spectra, and did not correct for irradiation effects. We therefore omit this measurement 
from the calculation. See Appendix~C.

\item {\it IP Peg. }
\scite{marsh88a} applied the light centres method to obtain $K_W$. \scite{martin89} found 
$K_R$ and $\Delta \phi_{1/2}$ which provides a direct spectroscopic determination of the 
mass and radius of the secondary star. 

\item {\it CN Ori. }
\scite{friend90a} measured $K_R$, and combined this with possibly unreliable values of 
$i$ and $K_W$ derived from a radial velocity study by \scite{mantel87}, the details of 
which are unpublished.

\item {\it UU Aqr. }
\scite{baptista94} used the method of \scite{wood89} to calculate the system parameters 
of this nova-like from photometry of the eclipses. Unfortunately the eclipse light curves 
do not have the shape usually required for the eclipse technique to be completely valid. 
We therefore group this mass determination with the possibly unreliable $K_W$ determinations.

\item {\it U Gem. }
The value of $K_R$ found by \scite{friend90a} was combined with $K_W$ measured by \scite{stover81c}
and $i$ estimated by \scite{smak76} from the location of the bright spot to calculate 
the masses. 

\item {\it BD Pav. }
$K_R$ and $V \sin i$ were measured by \scite{friend90a}, and have been combined with the measurement 
of $\Delta\phi_{1/2}$ (\pcite{harrop96} and references therein) to obtain the mass and radius of the 
secondary star. See Appendix~C.

\item {\it DQ Her. }
\scite{horne93} measured $K_R$ and $V \sin i$ and combined these with previous measurements of $K_W$ 
and $\Delta\phi_{1/2}$ to obtain a full set of system parameters, using a Monte Carlo simulation. 

\item {\it IX Vel. }
\scite{beuermann90} detected emission lines emanating from the secondary star in this bright, 
non-eclipsing nova-like, which effectively make it a double-lined binary. Using kinematic and 
geometric modelling, and using the Balmer line light curves to constrain $i$, they 
obtained the mass and radius of the secondary star.

\item {\it V895 Cen (=\,EUVE J1429\,+\,38.0). }
The mass and radius of the secondary have recently been determined by \scite{buckley98}. $K_R$ and 
$V \sin i$ were obtained using the Na{\small I} 8190\AA\ absorption doublet and $i$ 
was found from ellipsoidal variations.

\item {\it EX Dra (=\,HS 1804\,+\,6753).}
\scite{fiedler97} measured $K_W$ and $K_R$ and obtained $i$ from the geometry of the white dwarf 
and bright spot eclipses. However the radial velocity curve of the H$\alpha$ emission line is out of 
phase by $\sim 0.2$ and the value of $K_W$ determined is therefore unreliable. We have recomputed the 
system parameters using a Monte Carlo simulation with the values of $K_R$ and $V \sin i$ derived by 
\scite{billington96} and $\Delta\phi_{1/2}$ derived by \scite{fiedler97}. See Appendix~C.

\item {\it EM Cyg. }
This double-lined dwarf nova has glancing eclipses, and has had the radial velocities of both 
components measured by \scite{stover81a}. Unusually, the secondary star is more massive than 
the white dwarf. This means that EM~Cyg lies outside the ($M_2,q$) range of thermal and dynamical 
stability of mass transfer, assuming standard properties of Population I stars. These are, however, 
sensitive to metallicity, opacities and convection theory used to compute the models. The fact 
that EM~Cyg is observed to be in a moderate state of mass transfer means the star is probably 
not unstable.

\item {\it AC Cnc. }
Another eclipsing double-lined system \cite{schlegel84}, also with $q > 1$.

\item {\it V363 Aur (Lanning 10). }
\scite{schlegel86} obtained the mass and radius of the secondary star through measurements of 
$K_R$ and $K_W$ and estimated $i$ using $\Delta\phi_{1/2}$.

\item {\it BT Mon. }
\scite{smith98} measured $K_R$ from the weak secondary star absorption lines using skew mapping, 
$K_W$, $V \sin i$ and $\Delta\phi_{1/2}$ to obtain a full set of system parameters using a Monte
Carlo simulation.

\item {\it AE Aqr. }
\scite{casares96} detected absorption features from the secondary star and obtained measurements of $K_R$ 
and $V \sin i$. By modelling the way $V \sin i$ changes with orbital phase, they were able to constrain $i$.

\item {\it DX And. }
\scite{drew93} measured $K_R$ and $V \sin i$ and also estimated $K_W$. \scite{hilditch95} used 
ellipsoidal variations to provide an estimate of $i$. We have used this estimate with the 
measurements of Drew et al. to calculate the mass and radius of the secondary star. We 
prefer the $q$ determination of Drew et al. over that derived by \scite{bruch97} because Bruch et al. 
used a phase-shifted $K_W$ measurement from poorly wavelength calibrated data.

\item {\it V616 Mon (A0620-00).}
\scite{marsh94} measured $K_R$, $V \sin i$ and ellipsoidal variations in the equivalent width of the 
H$\alpha$ emission line. We have used their $1\sigma$ errors. A further constraint on $i$ is given by 
the grazing eclipses observed by \scite{haswell93}, giving $i \sim 70^\circ$. However, infra-red photometry 
by \scite{shahbaz94a} of the ellipsoidal variations give a lower inclination, $i\sim 40^\circ$, which 
implies a higher mass and radius, $M_2 \sim 0.6 M_\odot$ and $R_2 \sim 0.8 R_\odot$. 

\item {\it QZ Vul (GS 2000+25).}
\scite{harlaftis96} measured $V \sin i$ and $K_R$ and combined these measurements with a wide 
range of allowable values of $i$ to obtain an estimate of the mass of the secondary.

\item {\it GU Mus (Nova Mus 1991, GRS 1124-68).}
\scite{casares97} measured $K_R$ and $V \sin i$ from the secondary absorption lines. $i$ has been 
estimated from the ellipsoidal variations \cite{orosz96}. Using these parameters we have calculated 
the mass and radius of the secondary using a Monte Carlo simulation. See Appendix~C.

\item {\it V1033 Sco (Nova Sco 1994, GRO J1655-040).}
\scite{orosz97} measured $K_R$ spectroscopically and modelled the light curves, which contained 
substantial ellipsoidal variations, to obtain $q$ and $i$. 

\item {\it V404~Cyg (GS 2023+338)}
\scite{casares94a} measured $V \sin i$ and $K_R$ spectroscopically. \scite{shahbaz94b} modelled the  
ellipsoidal variations to obtain $i$ and hence the system parameters. See also \scite{shahbaz96d}.

\end{enumerate}

\section*{Appendix B}
\subsection*{Simple formulae for estimating the mass ratio and secondary star mass}

While the best method of calculating the masses of primary and secondary stars and the other 
system parameters in CVs and LMXBs is the Monte Carlo method (e.g. \pcite{smith98}), simple 
approximations can give fairly precise estimates of $q$ and $M_2$.

$K_R$ and $V \sin i$ are respectively given by
\begin{equation}
K_R = {{2 \pi}\over P} {a \over {(1+q)}} \sin i
\end{equation}
and
\begin{equation}
V \sin i = {{2 \pi}\over P} R_2 \sin i.
\end{equation}
Combining these two equations gives
\begin{equation}
{{R_2} \over a}{(1+q)} = {{V \sin i} \over {K_R}},
\end{equation}
which using equation~(2) yields a simple cubic formula for approximating $q$,
\begin{equation}
q(1+q)^2 = 9.6 \Bigl({{V \sin i} \over {K_R}}\Bigr)^3.
\end{equation}

Also, the density--period relation given by equation~(3) can be used in combination with 
equation~(16) to give a simple estimate of the mass of the secondary in terms of $V\sin i$ 
and $P$ if $i$ is known roughly

\begin{equation}
\Bigl({{M_2} \over{M_\odot}} \Bigr)= 0.042 \Bigl({{V \sin i} \over {100\,{\rm km}\,{\rm s}^{-1}}}\Bigr)^3 {{P({\rm hr})} \over {\sin^3 i}}.
\end{equation}

For eclipsing systems, $\sin^3 i$ can be approximated as 0.98 without introducing a significant error.
This then leads to the mass of the secondary being solely dependent on $V \sin i$ and $P$.
\begin{equation}
\Bigl({{M_2} \over{M_\odot}} \Bigr)= 0.043 \Bigl({{V \sin i} \over {100\,{\rm km}\,{\rm s}^{-1}}}\Bigr)^3 {P({\rm hr})}.
\end{equation}

\section*{Appendix C}
\subsection*{Results from Monte Carlo simulations}
A number of the secondary star masses and radii have been derived using Monte Carlo simulations
with values of $P$, $K_R$, $V \sin i$, $i$ and $\Delta \phi_{1/2}$ recovered from the literature.
The method used is similar to that described by \scite{smith98}. The resulting system parameters 
are given in Table~5.

\begin{table*}
\label{tab:monte}
\caption{Results from the Monte Carlo simulations. Parameters in bold face were those input to the Monte 
Carlo simulator.}
\begin{tabular}{cccccc} 
\hline
\multicolumn{1}{c}{} & 
\multicolumn{1}{c}{AM Her} &
\multicolumn{1}{c}{BD Pav} & 
\multicolumn{1}{c}{EX Dra} & 
\multicolumn{1}{c}{DX And} & 
\multicolumn{1}{c}{GU Mus} \\ 
\hline
$P\,$(hr) 		& {\bf 3.09}	 & {\bf 4.30}		& {\bf5.04}		& {\bf10.60}	 	& {\bf10.38}	\\
$K_R$ (km\,s$^{-1}$)	& {\bf 179$\pm$2}& {\bf 278$\pm$4}	& {\bf 210$\pm$14}	& {\bf 105.8$\pm$3.8} 	& {\bf 420.8$\pm$6.3}\\
$K_W$ (km\,s$^{-1}$) 	& $115\pm18$	 & $123\pm17$		& $176\pm19$		& $103\pm9$	 	& $56\pm17$	\\
$V\sin i$ (km\,s$^{-1}$) & {\bf 100$\pm$10} & {\bf 125$\pm$10}	& {\bf 140$\pm$10} 	& {\bf 79$\pm$5} 	& {\bf 106$\pm$13}	\\
$i$	 		& {\bf 52$\pm$5} & $73.4\pm0.9$		& $82.1\pm2.0$		& {\bf 49$\pm$4} 	& {\bf 59.5$\pm$5.5}	\\
$\Delta \phi_{1/2}$ 	&	 --	 & {\bf 0.040$\pm$0.006}& {\bf 0.1103$\pm$0.0001} & 	-- 	 	&  	-- 	\\
$M_1$ ($M_\odot$) 	& $0.44\pm0.11$	 & $0.95\pm0.10$	& $0.70\pm0.10$		& $0.51\pm0.12$	 	& $6.98\pm1.45$\\
$M_2$ ($M_\odot$) 	& $0.29\pm0.10$	 & $0.43\pm0.10$	& $0.59\pm0.12$		& $0.50\pm0.14$	 	& $0.94\pm0.40$	\\
$R_2$ ($R_\odot$) 	& $0.33\pm0.04$	 & $0.46\pm0.04$	& $0.59\pm0.04$		& $0.92\pm0.08$	 	& $1.06\pm0.15$	 \\
$q$ 			& $0.64\pm0.10$	 & $0.44\pm0.06$	& $0.85\pm0.12$		& $0.98\pm0.10$	 	& $0.13\pm0.04$ \\
$a$ ($R_\odot$)		& $0.96\pm0.09$	 & $1.48\pm0.05$	& $1.58\pm0.08$	 	& $2.44\pm0.19$	 	& $4.77\pm0.34$ \\
\hline
\end{tabular}
\end{table*}

\end{document}